\documentclass{aa}
\usepackage{epsfig}
\usepackage{psfig}
\usepackage{graphicx}
\usepackage{txfonts}

\usepackage{lscape}

\newcommand{\psrA}{PSR J1824$-$2452A}
\newcommand{\psrH}{PSR J1824$-$2452H}
\newcommand{\CXO}{{\sl CXO}}

\titlerunning{A search for X-ray counterparts of the millisecond pulsars 
in the globular cluster M28}

\begin{document}
\title{A search for X-ray counterparts of the millisecond pulsars
in the globular cluster M28 (NGC 6626)}

\author{W.~Becker \and C.Y.~Hui}   
\institute{Max-Planck Institut f\"ur Extraterrestrische Physik, 
          Giessenbachstrasse 1, 85741 Garching bei M\"unchen, Germany}
\date{submitted on May, 1th, 2007}

\abstract{A recent radio survey of globular clusters has increased the
 number of millisecond pulsars drastically. M28 is now the globular cluster 
 with the third largest population of known pulsars, after Terzan 5 and 47\,Tuc. 
 This prompted us to revisit the archival Chandra data on M28 to evaluate 
 whether the newly discovered millisecond pulsars find a counterpart 
 among the various X-ray sources detected in M28 previously. The
 radio position of \psrH\ is found to be in agreement with the position of    
 CXC 182431-245217 while some faint unresolved X-ray emission near to the
 center of M28 is found to be coincident with the millisecond pulsars 
 PSR J1824-2452G, J1824-2452J, J1824-2452I and J1824-2452E. 

}

\maketitle

\section{Introduction}
 About 10\% of the 1765 known radio pulsars are millisecond pulsars (MSPs). The 
 majority of them, currently 134, are located within 24 globular
 clusters\footnote{http://www2.naic.edu/~pfreire/GCpsr.html} (GCs) which, because
 of their extreme stellar core density, lead to dynamical interactions that apparently 
 play a significant role in the formation of MSPs. The first MSP discovered in a globular 
 cluster was \psrA\ in M28 (Lyne et al.~1987). Its inferred pulsar parameters 
 make it the youngest ($P/2\dot{P}=3.0 \times 10^7$ yrs) and most powerful ($\dot{E}=2.24
 \times 10^{36}$ erg s$^{-1}$) pulsar among all known MSPs. A recently performed deep 
 radio survey of M28 which was designed to search for more cluster pulsars led to the 
 discovery of ten new MSPs (B\'egin 2006). 
 Together with the previously known \psrA, four of the M28 MSPs appear to be solitary. 
 The others are in binaries. Two pulsars are found to be in highly eccentric 
 binary systems with eccentricities of $e=0.847$ and 0.776, respectively, and companion 
 masses of at least 0.26 $M_\odot$ and 0.38 $M_\odot$ (B\'egin 2006). M28 is now the 
 globular cluster with the third largest population of known MSPs pulsars, after 
 Terzan 5 and 47\,Tucanae (Camilo et al.~2000, Ransom et al.~2005, Stairs et al.~2006).
 
 In this letter we report on a re-analysis of Chandra ACIS-S data from the globular
 cluster M28. It was performed to search for X-ray counterparts of the newly 
 discovered MSPs. The re-analysis builds partly on the analysis of the M28 Chandra 
 data which was published by Becker et al.~(2003).
 
\section{Observations and Data Analysis} 
 M28 was observed by the Chandra ACIS-S three times for approximately equal 
 observing intervals of about 13 ksec between July and September 2002. These 
 observations were scheduled so as to be sensitive to time variability on time 
 scales up to weeks.  The observations were made using 3 of the \CXO\ Advanced 
 CCD Imaging Spectrometer (ACIS) CCDs (S2,3,4) in the faint timed exposure mode 
 with a frame time of 3.241\,s. Standard {\sl Chandra} data processing 
 has applied aspect corrections and compensated for spacecraft dither. Level~2 
 event lists were used in our analyses. Events in pulse invariant channels 
 corresponding to $\approx 0.2$ to 8.0 keV were selected for the purpose of finding 
 sources. Increased background corrupted a small portion of the third data set 
 reducing its effective exposure time from 14.1 ksec to 11.4 ksec although no 
 results were impacted by the increased background.

 The optical center of the cluster at $\alpha_{2000} = 18^{\rm h}\,24^{\rm m}\,32\fs89$
 and $\delta_{2000} = -24\degr\, 52\arcmin\, 11\farcs4$ (Shawl and White 1986) was
 positioned 1\arcmin\ off-axis to the nominal aim point on the back-illuminated CCD,
 ACIS-S3, in all 3 observations. A circular region with 3\farcm1-radius, corresponding
 to twice the half-mass radius of M28, centered at the optical center was extracted
 from each data set for analysis.  No correction for exposure was deemed necessary
 because the small region of interest lies far from the edges of the S3 chip.

 Applying a wavelet source detection algorithm, the X-ray position of \psrA\ 
 was measured separately using the three data sets and the
 merged data. The set-averaged position is the same as that derived using the merged 
 data set and was found to be in agreement with the pulsar's radio position. The result
 is summarized in Table 1. The root-mean-square (rms) uncertainty in the pulsar position, 
 based on the 3 pointings, is 0\farcs042 in right ascension and 0\farcs029 in declination.  
 The radio position and proper motion of the pulsar, as measured by Rutledge et al.\ (2003), 
 places the pulsar at the time of the observation only $\Delta_{\alpha}=$ 0\farcs083, 
 $\Delta_{\delta}=-0\farcs042$ away from the best-estimated X-ray position.  In what follows 
 the X-ray positions of all sources have been adjusted to remove this offset. 

\begin{table}[h!!!!!] \centering
 \caption{\psrA\ Positions (J2000)\label{psr_positions}}
 \begin{tabular}{cccc}
 \hline\hline
   Date & Position \\ \hline\\[-1.5ex]
   2002 July 4 &  18 24 32.015 &  $-$24 52 10.81 \\
   2002 Aug  8 &  18 24 32.016 &  $-$24 52 10.76 \\
   2002 Sep  9 &  18 24 32.009 &  $-$24 52 10.83 \\
   average     &  18 24 32.013 &  $-$24 52 10.80 \\
   rms (arcsec)& 0.042         &  0.029        \\
   merged data & 18 24 32.013  & $-$24 52 10.80  \\
   radio (8/02)& 18 24 32.008  & $-$24 52 10.76  \\\hline\\[-1.5ex]
 \end{tabular}
\end{table}

 In addition to the 37.6 ksec of ACIS-S data we reanalyzed about 100 ksec of HRC
 data which were available from M28 in the Chandra archive. Part of this data 
 were taken in 2002 October 11.~and were used by Rutledge et al.~(2003) to 
 investigate the temporal X-ray emission properties of \psrA. The second half 
 of the HRC-S data was taken more recently, in 2006 May 27. \& 28., for the 
 purpose of Chandra on-board clock calibration.  However, both HRC data sets 
 were taken in {\em timing mode} and hence suffer from very high background 
 as in this mode all photons which are registered in the detector are 
 transmitted to the ground to allow a correction for the detectors timing 
 bug (cf.~Tenant et al.~2001). As far as the detection of faint sources is 
 concerned the available ACIS-S data supersede the HRC data in sensitivity, 
 albeit more than 60\% shorter in exposure time. We therefore did not 
 further consider the HRC data for the search of X-ray counterparts of 
 the newly discovered MSPs.

 Amongst the many interesting results we obtained from the ACIS-S observation 
 and which were reported already in detail in Becker et al.~(2003), this data 
 detected 46 X-ray sources in a field of 4 arcmin near to the pulsar \psrA. 12 
 of these sources are located within the 14.4 arcsec cluster core radius. The 
 properties of these sources along a detailed spectral analysis for the brightest 
 among them was published in Becker et al.~(2003) so that we can omit to repeat all
 details of our previous analysis here again. To briefly summarize few basic 
 findings in the following, though, might be convenient for the reader. 
 
 The brightest source in the ACIS-S data (\#26 in Table 3 of Becker et al.~2003), 
 which is also the one with the softest spectrum, was identified as a candidate 
 LMXB in quiescence, whereas all the other sources in the field turned out to 
 have rather hard X-ray spectra. Source \#19 (cf.~Table 3 of Becker et al.~2003
 and Figure 1 below), which emits the hardest X-rays, is the millisecond pulsar 
 \psrA. Several of the other X-ray sources seen in M28 were found to show 
 variability on time scales of hours to weeks and are still unidentified. 

 The superior $\sim 1$ arcsec angular resolution of Chandra not only allowed 
 us to resolve \psrA\ from nearby sources which in the ROSAT HRI were 
 only seen as diffuse unresolved emission, it furthermore provided the first 
 uncontaminated, phase-averaged, spectrum from the brightest among all
 millisecond pulsars. This spectrum was found to be best described by a power 
 law with photon index $1.2^{+0.15}_{-0.13}$. \psrA\ thus has a very hard 
 X-ray spectrum which made it to detect the pulsar up to $\sim 20$ keV with 
 RXTE (Mineo et al.~2004). Exciting to note and most interesting, however,
 is the evidence of an emission line feature centered at $\sim 3.3$ keV in the pulsar
 spectrum (cf.~Fig.~2 in Becker et al.~2003). This line feature can be interpreted as
 cyclotron emission from a corona above the pulsar's polar cap if the magnetic field
 is different from a centered dipole configuration. The significance of the 
 feature, however, is at the edge of detectability which prevents any final 
 conclusion and clearly calls for confirmation in a further deeper observation.

 Figure 1 shows the central region of M28 as seen by Chandra ACIS-S3 between 
 July and September 2002. The image was created with a spatial binning
 of 0.5 arcsec. Positions of \psrA\ and nine newly discovered millisecond 
 pulsars\footnote{The position of J1824$-$2452K is not avaliable in B\'egin (2006)}
 are indicated (cf.~Table 2). Inspecting the data for possible emission from 
 the newly discovered MSPs shows that there is some faint unresolved 
 X-ray emission near to the center of M28 (cf.~Figure 1) which is coincident with 
 the locations of four of the new MSPs: PSR J1824-2452G, J1824-2452J, 
 J1824-2452I and J1824-2452E. The position of the eclipsing binary millisecond pulsar 
 PSR J1824$-$2452H reported by B\'egin (2006) is found to be $\sim 0.2$ arcmin north from 
 the position of CXC 182431-245217 which is source \#18 reported in Becker et al.~(2003). 
 More accurate radio timing solutions, however, had found that the position of 
 PSR J1824$-$2452H is indeed in agreement with CXC 182431-245217 (Ransom 2007, priv.~com.).  
 The identification of this X-ray source as a pulsar is further supported by the 
 similarity of its hardness ratio with that of \psrA\ and the lack of long-term 
 variability (cf.~Table 3 in Becker et al. 2003). By measuring the counts within 
 a circle of 2 arcsec (encircled energy $\sim 95$\%) at the MSP source positions, 
 we have determined the counting rates and upper limits for the 10 out of 11 pulsars 
 in M28. The results are listed in Table 2 together with the a summary of the 
 millisecond pulsar parameters from B\'egin (2006). 

\begin{center}
\begin{figure*}[t!!!!!!]
  \hspace{-1.5cm}{\psfig{figure=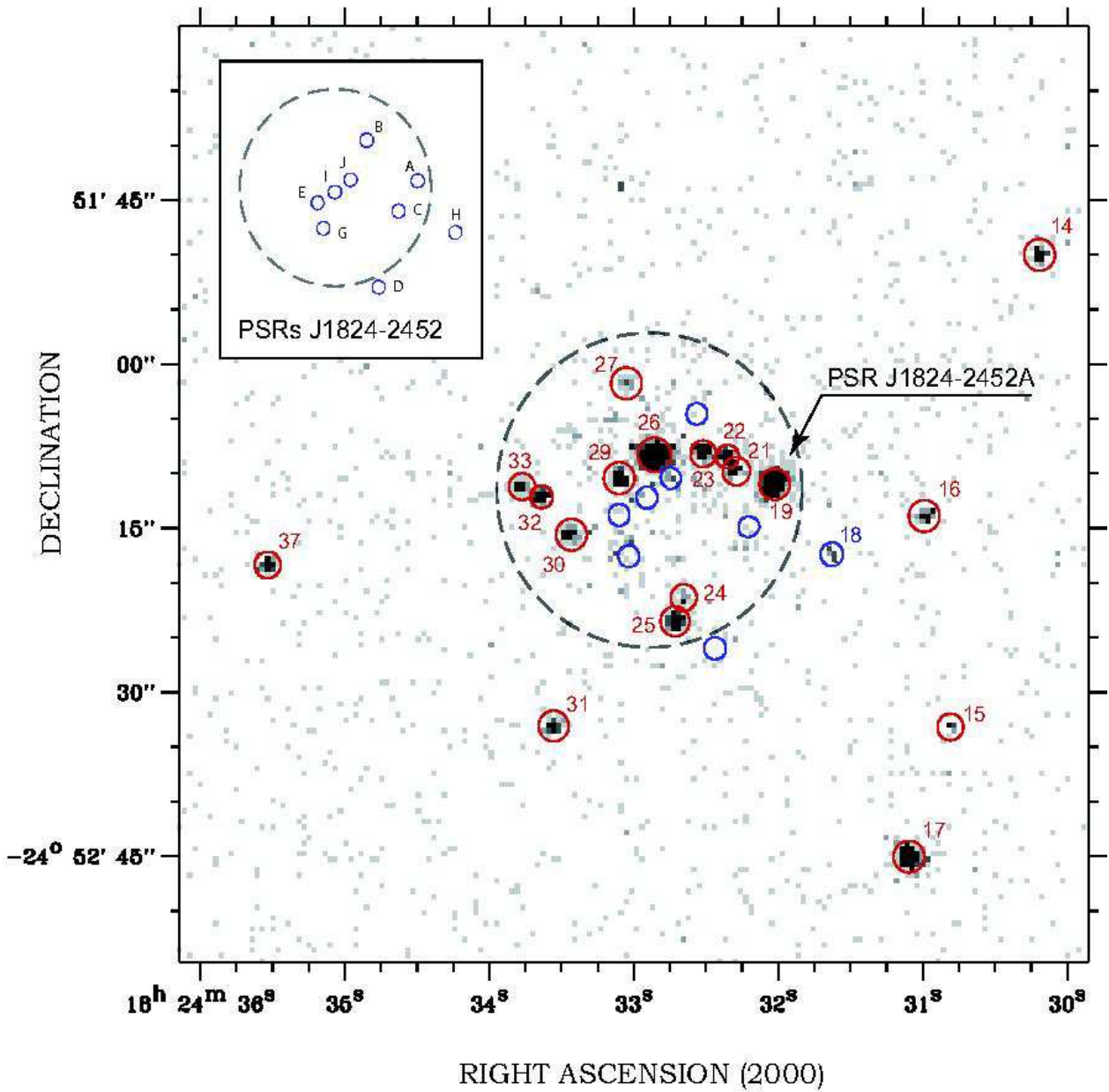,width=18cm,clip=}}
  \caption[]{\small {\sl Chandra} ACIS-S3 image of the central region of M28.
  The positions of the recently discovered new ms-pulsars are indicated by
  blue circles. Most are located within or near to the core-radius (dashed
  circle). One of the new ms-pulsars is placed outside the field shown in
  this plot. The binary millisecond pulsar PSR J1824$-$2452H is in agreement
  with source \#18 reported in Becker et al.~(2003). There is faint (only partly
  resolved) emission near to the center of the cluster which is in agreement
  with the location of the new pulsars PSR J1824$-$2452J, J1824$-$2452I and
  J1824$-$2452E. The inset helps to locate the different MSPs.}
\end{figure*}
\end{center}

 Because of the low ACIS-S instrument background, the counts obtained from the
 location of the MSPs J1824-2452G, J1824-2452I and J1824-2452J suggest that there
 is significant emission from these sources, albeit unresolved and at the edge of 
 sensitivity in the available data. This suggests, however, that in a deeper Chandra 
 ACIS-S observation it will be possible to resolve and detect the emission from all 
 these new pulsars and to obtain spectral information from them. The pulsar
 PSR J1824$-$2452H (respectively CXC 182431-245217) is detected with a 
 signal-to-noise ratio of 2.88. Determing its counting rate yields $(2.9\pm 0.9) 
 10^{-5}$ cts/s, corresponding to an $0.2-8.0$ keV X-ray luminosity of 
 $1.2\times 10^{30}$ erg/s for an assumed power-law spectrum with a photon-index
 of 2, a source distance of 5.5 kpc and a column absorption of $0.18\times
 10^{22}$\,cm$^{-2}$. Albeit this is only a rough luminosity estimate, the
 X-ray conversion concluded from it is only $3.6 \times 10^{-5}$.

\begin{table*}
\centering
 \caption{Basic properties of millisecond pulsars located in the globular cluster 
  M28 and their counting rates in Chandra ACIS-S data.\newline The position of J1824-2452H is 
  the Chandra X-ray position while for the other pulsars it is the radio timing 
  position from B\'egin (2006).}
 \begin{tabular}{c c c c c c c c}\hline\hline\\[-1ex]
   Name      &     Ra          &      Dec      &      P        &     $\dot{E}$     &   $B_\perp$   &  ACIS-S     &    rate    \\[1ex]
    {}       &    H:M:S        &      D:M:S    &   $10^{-3}$ s &   $10^{34}$ erg/s &   $10^{9}$ G  & net counts  &    cts/s   \\[1ex]\hline\\[-1.5ex]
 J1824-2452A &  18 24 32.007 &   -24 52 10.49  &   3.0543      &      22.243       &      {}       &     1100    &   2.63e-2  \\
 J1824-2452B &  18 24 32.545 &   -24 52 04.29  &   6.5466      &       {}          &    $<$ 0.4    &      3      &   7.89e-5  \\
 J1824-2452E &  18 24 33.089 &   -24 52 13.57  &   5.4191      &       {}          &    $<$ 0.8    &      5      &   1.32e-4  \\
 J1824-2452F &  18 24 31.812 &   -24 49 25.03  &   2.4511      &     2.5198        &    $<$ 0.5    &      1      &   2.63e-5  \\[1ex]

\multicolumn{6}{c}{Binary pulsars:}\\[1ex]

 J1824-2452G &  18 24 33.025 &   -24 52 17.32  &   5.90905    &      3.4228        &  $<$ 1.6      &     17      &   4.47e-4  \\
 J1824-2452H &  18 24 31.591 &   -24 52 17.49  &   4.62941    &      3.2586        &  $<$ 1.1      &     11      &   2.89e-4  \\
 J1824-2452I &  18 24 32.9   &   -24 52 12.00  &   3.93180    &        {}          &   {}          &     12      &   3.16e-4  \\
 J1824-2452J &  18 24 32.733 &   -24 52 10.18  &   4.03968    &        {}          &  $<$ 0.6      &     22      &   5.78e-4  \\[1ex]

\multicolumn{6}{c}{Eccentric binary pulsars:}\\[1ex]

 J1824-2452C &  18 24 32.192 &   -24 52 14.66  &   4.15828   &       9.3416        &  $<$ 1.2      &      3       &   7.89e-5  \\
 J1824-2452D &  18 24 32.422 &   -24 52 25.90  &   79.8354   &       7.6183        &  $\sim 91.0$  &      2       &   5.26e-5  \\\hline\hline
\end{tabular}
\end{table*}

\section{Discussion}

Since the {\sl Einstein} era it has been clear that globular cluster contain various populations
of X-ray sources of very different luminosities (Hertz \& Grindlay 1983). The stronger
sources ($L_{x} \approx 10^{36}-10^{38}$ {${\rm erg}\, {\rm s}^{-1}$}) were seen to
exhibit X-ray bursts which led to their identification as low-mass X-ray binaries (LMXBs).
The nature of the weaker sources, with $L_x \le 3\!\times \! 10^{34}$ {${\rm erg}\, {\rm s}^{-1}$},
however, was more open to discussion (e.g., Cool et al.~1993; Johnston \& Verbunt 1996).
Although many weak X-ray sources were detected in globulars by ROSAT (Johnston \& Verbunt
1996; Verbunt 2001), their identification has been difficult due to low photon statistics
and strong source confusion in the crowded globular cluster fields. It was therefore clear
that Chandra with its sub-arcsecond angular resolution would contribute tremendously to the
investigation of globular clusters. The results which have been published on this subject
so far has shown that these expectations were justified. Among the results we obtained on 
M28 (Becker et al.~2003) important work has been done on Terzan 5 (Heinke et al.~2006) and 
of 47~Tuc = NGC 104. From the latter, Grindlay et al.~(2001) reported the detection of 108 
sources within a region corresponding to about 5 times the 47~Tuc core radius. Nineteen of 
the soft/faint sources were found to be coincident with radio-detected millisecond pulsars 
(Bogdanov et al.~2006) and Grindlay et al.~(2001a, 2002) concluded that more than 50 percent 
of all the unidentified sources in 47~Tuc are MSPs. This conclusion is in concert with 
theoretical estimates on the formation scenarios of short-period (binary) pulsars in globular 
clusters (e.g.~Rasio, Pfahl \& Rappaport 2000).

Some of the unresolved excess of X-ray emission from within the central part of M28
that we reported in Becker et al.~(2003) is likely  due to point sources that were
just below the 38-ksec sensitivity limit. The recent discovery of the new MSPs, which
are partly in positional agreement with this excess emission, strongly support the
conjecture. We also know from Chandra observations of 47 Tuc that the soft emission
from millisecond pulsars, like those found in that cluster, tend to be below the
sensitivity limit of the previous Chandra M28 observation because of the higher
absorption towards M28. Hence it is quite likely that a deeper exposure will reveal
more, faint, X-ray sources which might turn out to be millisecond pulsars as well.
Apart from resolving the diffuse emission, a second deeper observation will add
significant information to those faint sources for which a detailed spectral modeling
was precluded by limited photon statistics in the existing 38-ksec ACIS-S data.

As all these MSPs in M28 are virtually located at the same distance from us detecting
them will allow one to investigate their relative X-ray efficiency, $L_x/\dot{E}$, 
unbiased by distance uncertainties. Grindlay et al.~(2002) found that the dependence 
of $L_x$ on $\dot{E}\,$ for 47 Tuc MSPs may be $L_x \propto \dot{E}^{0.5}$, i.e.~significant 
flatter than the $L_x/\dot{E}=10^{-3}$ observed for pulsars located in the galactic 
plane (cf.~Becker \& Tr\"umper 1997). The 47 Tuc MSPs were found to be consistent 
with the X-radiation being emitted from heated polar caps. Whether this will be the 
dominating X-ray emission process of the newly discovered MSPs in M28 and whether 
their X-ray efficiency will be consistent with $L_x \propto \dot{E}^{0.5}$ or with 
$L_x/\dot{E}=10^{-3}$ will be a question one can address with data of higher photon 
statistics.

Several of the other sources we detected previously in M28 exhibit variability on
time scales of weeks. Our M28 X-ray survey also found a large number of objects that
were bright in only a single observation.  Some of these latter objects were bright
because they flared during the observation (i.e., not only were they bright in one
observation, they were bright for only part of the observation). We suspect that
these sources are flare stars (RS CVn or BY Dra). There were a few additional
sources seen in the HRC observation of M28 by Rutledge et al.~(2004), performed
only a few weeks after the ACIS-S observations. These sources either turned on at,
or in between, the two observations. As long as each star is only seen to flare in
one observation then all that we know is that the total population is larger than
the number that we have seen in the previous 38-ksec observation. However, if one 
sees the same flare stars more than once it will allow one to better estimate the 
total source population.

\end{document}